\documentclass[sn-mathphys-num]{sn-jnl}% Math and Physical Sciences Numbered Reference Style

\usepackage{graphicx}%
\usepackage{multirow}%
\usepackage{amsmath,amssymb,amsfonts}%
\usepackage{amsthm}%
\usepackage{mathrsfs}%
\usepackage[title]{appendix}%
\usepackage{xcolor}%
\usepackage{textcomp}%
\usepackage{manyfoot}%
\usepackage{booktabs}%
\usepackage{algorithm}%
\usepackage{algorithmicx}%
\usepackage{algpseudocode}%
\usepackage{listings}%
\begin{document}

\title{Deep Learning Enhanced Quantum Holography with Undetected Photons}

\author[1]{\sur{Weiru} \fnm{Fan}} 
\equalcont{These authors contributed equally to this work.}
\author[1]{\sur{Gewei} \fnm{Qian}}
\equalcont{These authors contributed equally to this work.}
\author[2]{\sur{Yutong} \fnm{Wang}} 
\author*[1]{\sur{Chen-Ran} \fnm{Xu}} \email{crxu@zju.edu.cn}
\author[3]{\sur{Ziyang} \fnm{Chen}}
\author[4]{\sur{Xun} \fnm{Liu}}
\author[4]{\sur{Wei} \fnm{Li}}
\author[5]{\sur{Xu} \fnm{Liu}}
\author[2]{\sur{Feng} \fnm{Liu}}
\author*[1]{\sur{Xingqi} \fnm{Xu}} \email{xuxingqi@zju.edu.cn}
\author[1,5,6]{\sur{Da-Wei} \fnm{Wang}}
\author*[7]{\sur{Vladislav V.} \fnm{Yakovlev}}\email{yakovlev@tamu.edu}

\affil[1]{\orgdiv{Zhejiang Province Key Laboratory of Quantum Technology and Device, School of Physics, and State Key Laboratory for Extreme Photonics and Instrumentation}, \orgname{Zhejiang University}, \orgaddress{\city{Hangzhou}, \postcode{310027}, \state{Zhejiang Province}, \country{China}}}
\affil[2]{\orgdiv{College of Information Science and Electronic Engineering}, \orgname{Zhejiang University}, \orgaddress{\city{Hangzhou}, \postcode{310027}, \state{Zhejiang Province}, \country{China}}}
\affil[3]{\orgdiv{College of Information Science and Engineering, Fujian Key Laboratory of Light Propagation and Transformation}, \orgname{Huaqiao University}, \orgaddress{\city{Xiamen}, \postcode{361021}, \state{Fujian Province}, \country{China}}}
\affil[4]{\orgdiv{Beijing Institute of Space and Electricity}, \orgname{China Academy of Space Technology}, \orgaddress{\city{Beijing}, \postcode{100094}, \country{China}}}
\affil[5]{\orgdiv{College of Optical Science and Engineering}, \orgname{Zhejiang University}, \orgaddress{\city{Hangzhou}, \postcode{310027}, \state{Zhejiang Province}, \country{China}}}
\affil[6]{\orgdiv{Hefei National Laboratory}, \orgaddress{\city{Hefei}, \postcode{230088}, \state{Anhui province}, \country{China}}}
\affil[7]{\orgdiv{Department of Biomedical Engineering}, \orgname{Texas A\&M University}, \orgaddress{\city{College Station}, \postcode{77843}, \state{TX}, \country{USA}}}

\abstract{Holography is an essential technique of generating three-dimensional images. Recently, quantum holography with undetected photons (QHUP) has emerged as a groundbreaking method capable of capturing complex amplitude images. Despite its potential, the practical application of QHUP has been limited by susceptibility to phase disturbances, low interference visibility, and limited spatial resolution. Deep learning, recognized for its ability in processing complex data, holds significant promise in addressing these challenges. In this report, we present an ample advancement in QHUP achieved by harnessing the power of deep learning to extract images from single-shot holograms, resulting in vastly reduced noise and distortion, alongside a notable enhancement in spatial resolution. The proposed and demonstrated deep learning QHUP (DL-QHUP) methodology offers a transformative solution by delivering high-speed imaging, improved spatial resolution, and superior noise resilience, making it suitable for diverse applications across an array of research fields stretching from biomedical imaging to remote sensing. DL-QHUP signifies a crucial leap forward in the realm of holography, demonstrating its immense potential to revolutionize imaging capabilities and pave the way for advancements in various scientific disciplines. The integration of DL-QHUP promises to unlock new possibilities in imaging applications, transcending existing limitations and offering unparalleled performance in challenging environments.}


\maketitle

\section*{Introduction}
Holography records and reconstructs the amplitude and phase of light scattered from objects by interference measurements \cite{gabor1948new}. Holography-based imaging offers realistic three-dimensional visualization, and has versatile applications in entertainment, scientific research, data encryption and biomedical engineering \cite{popescu2011quantitative,javidi2000securing,qu2020reprogrammable}. Entanglement can be used to enhance the spatial resolution \cite{defienne2022pixel} and robustness of holographic imaging \cite{defienne2021polarization}. However, such quantum holographic imaging based on joint measurement of the two entangled photons requires long acquisition time and sophisticated detection procedure, thereby hindering its applications. Quantum holography with undetected photons (QHUP) based on quantum induced coherence \cite{zou1991induced} only measures the intensity interference fringes of one of the two entangled light beams, which increases the imaging speed by more than 1000 times compared with the protocols using joint measurements \cite{black2023quantum}. More importantly, QHUP enables the mid-infrared and terahertz imaging with visible or near-infrared light detectors \cite{hochrainer2022quantum,kutas2021quantum}, which revolutionizes the fields such as optical imaging \cite{lemos2014quantum,gilaberte2021video,kviatkovsky2020microscopy,topfer2022quantum}, infrared spectroscopy \cite{kutas2021quantum,kalashnikov2016infrared,lindner2022accurate,lee2020molecular}, and optical coherence tomography \cite{vanselow2020frequency,paterova2018tunable,valles2018optical}. \par

However, compared to the dual-beam interferometry, QHUP relies on the phase stability of the pump, signal and idler light beams, and their precise mode overlap in the nonlinear crystal, which is prone to mechanical and optical disturbances. Such requirements pose obstacles to extracting amplitude-phase images from holograms with phase-shifted scanning~\cite{schnars2015digital,hariharan1987digital,zuo2022deep}. In addition, QHUP has a lower spatial resolution than its classical counterpart due to the requirement on the transverse phase matching in the crystal with a limited aperture~\cite{fuenzalida2022resolution,vega2022fundamental}. The photons used in QHUP have a shorter coherence length compared to those used in classical holography, such that the thickness variations in real samples can result in low interference visibility and deteriorated holograms when the mismatch between the light paths exceeds the coherence length \cite{qian2023quantum}. To reconstruct the images from such holograms, an elaborate and resource-consuming data processing is necessary. Deep learning which has been widely used in imaging~\cite{rivenson2017deep,li2018deep,durand2018machine}, spectroscopy \cite{ghosh2019deep,ho2019rapid} and adaptive optics \cite{hampson2021adaptive,feng2023neuws}, can solve the problems in QHUP by excavating complex data structure and correlation \cite{shimobaba2022deep}.\par

In this article, we develop an approach of deep learning enhanced quantum holography with undetected photons (DL-QHUP) which allows fast and accurate reconstruction of images from the one-shot hologram with low interference visibility. The key component of DL-QHUP is a convolution-based deep learning model with the state-of-the-art modules to extract information and reconstruct images layer by layer \cite{lecun2015deep}. By implementing the training of the model, DL-QHUP can automatically model the physical process and the noise of the inverse problem of holographic imaging to find the correspondence between the holograms and object images, enabling a robust imaging paradigm. By exploiting the pixel-to-pixel correlation of holograms, we find that DL-QHUP is capable of improving the spatial resolution and faithfully imaging for real objects. Furthermore, combined with the inherent background light resilience of QHUP, DL-QHUP exhibits an exceptional robustness to internal phase instability and external stray light noise. Merited by these advantages, our DL-QHUP offers a promising and reliable platform to realize one-shot video-rate bioimaging, industrial monitoring of materials and other complex application scenarios.\par

\section*{Results}

\begin{figure}[htb]%
\centering
\includegraphics[width=0.92\textwidth]{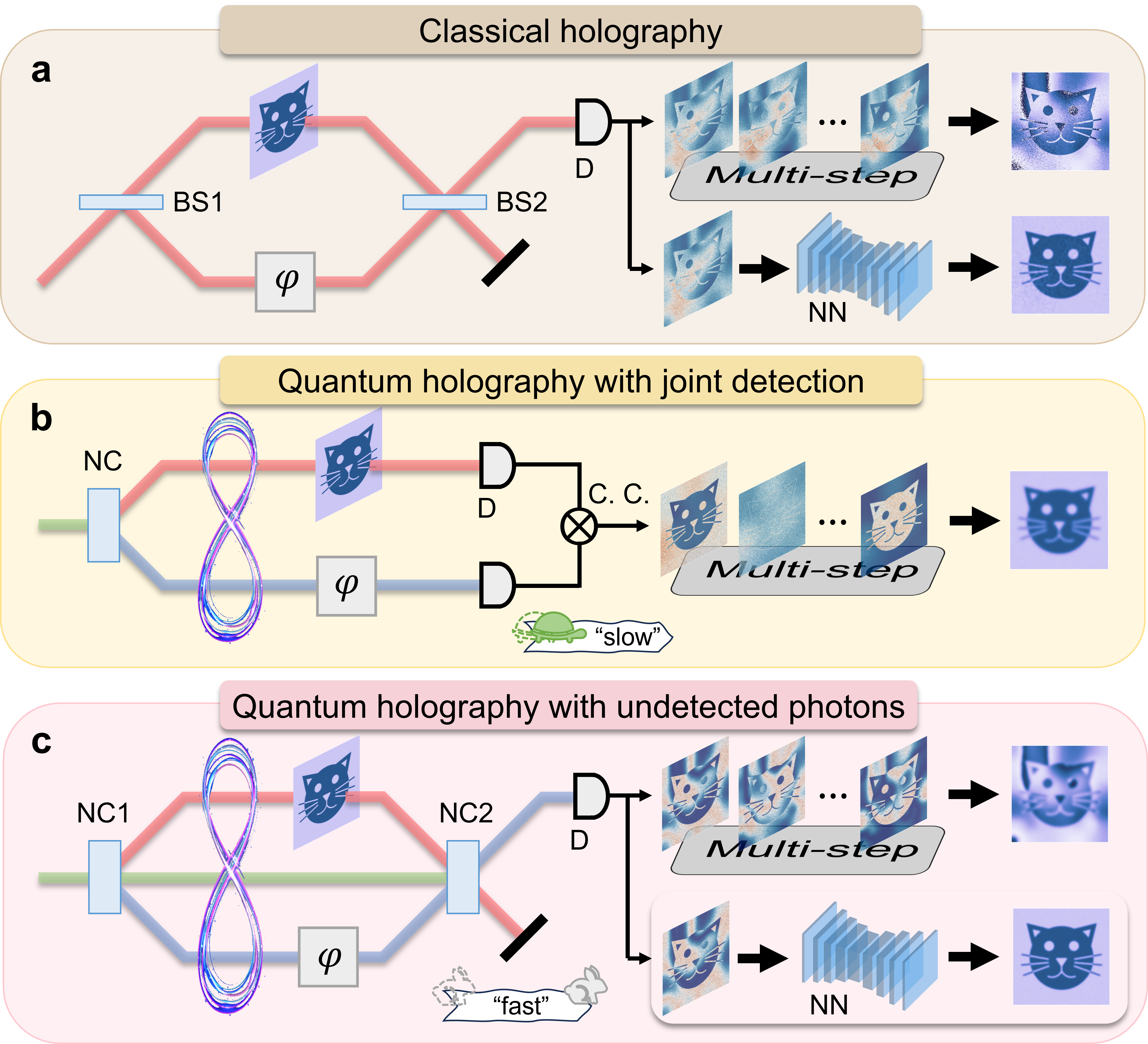}
\caption{\textbf{Comparison between different holographic imaging modalities.} (a) Classical holography based on dual-beam interference. Object information is recovered either via multi-step phase shifts which are vulnerable to phase disturbances or via neural network (NN) which enables the one-shot holography. (b) Quantum holography with joint detection. The complex amplitude of the object is encoded into the quantum state of light, offering strong phase-noise resistance but requiring long integration time. (c) Quantum holography with undetected photons (QHUP). QHUP records the complex amplitude of the object through the fast intensity measurements. Conventional QHUP retrieves images by multi-step phase shifts, susceptible to phase disturbances. DL-QHUP enables one-shot and robust holographic imaging. BS is beam splitter; D is detector; NN is neural network; NC is nonlinear crystal; C.C. is coincidence counting; $\varphi$ is phase shifter. See the comparison between different holographic modalities in supplememtary material Table. 1.}\label{fig1}
\end{figure}

The comparison between classical holography, quantum holography with joint detection and QHUP is illustrated in Fig. \ref{fig1}. Classical holography is based on a classical interferometer, where the illuminating arm records the complex amplitudes of the object and the reference arm introduces the phase shift to retrieve the image of the object. The phase noise accumulated in the single measurement and phase-shifted operation leads to a deteriorated image. Quantum holography enabled by the quantum state of light stores the complex amplitude of the object in the orthogonal polarization states of entangled photons, and is subsequently retrieved by crossed polarization analyzers and multi-step phase-shifted method \cite{defienne2021polarization}. Without requiring the spatial overlap of two beams, such quantum holography is immune to phase disorders arising from light scattering and mechanical vibrations of experimental set-up. Since the holographic information is encoded in second-order coherence, coincidence counting between the two entangled photons is necessary, which results in the long acquisition time on the order of hours.\par

Recently, the physical concept of quantum induced coherence has been used to realize holographic imaging, i.e., QHUP (shown in Fig. \ref{fig1}c, see Fig. S1 for the set-up in details). QHUP records the complex amplitude of the object by intensity measurements without joint detection, which relaxes the requirement of single-photon detectors and speeds up imaging. In the implementation of QHUP, a continuous-wave pump laser is sent into a nonlinear crystal to produce nondegenerate entangled photons, i.e., signal and idler photons, through spontaneous parametric down-conversion (SPDC). The entangled photons, along with the pump light, are then reflected to the crystal for a second SPDC. The object is placed in the path of idler photons, while the difference of distance between signal and idler photons is adjusted by introducing an optical delay of signal photons. When the distance difference is within the coherence length of the entangled photons, the interference pattern of signal photons can be observed by a camera. As a result of quantum induced coherence, the pathway information of signal photons is erased by the idler photons reflected by the object \cite{qian2023quantum}, such that the object information carried by the idler photons can be obtained through the detection of the signal photons.\par

The intensity distribution of signal photons on the camera can be described as \cite{gilaberte2021video,qian2023quantum}, 
\begin{equation}
    I_s \propto r_p^2+r_s^2+2r_pr_sr_i\left|\gamma\left(\Delta\tau\right)\right|\cos{\left(\varphi_s+\varphi_i-\varphi_p+\Delta\phi\right)},
\end{equation}
where the subscripts $s$, $i$ and $p$ represent the signal, idler, and pump photons, respectively, while $r$ is amplitude reflectivity and $\varphi$ is the phase in space. 
The envelope function, $\gamma$, is inversely proportional to the time delay $\left(\Delta\tau\right)$ between the signal and idler photons. We assume that $\gamma$ takes the form $\gamma\left(\Delta\tau\right) = \exp{\left[-\left(\Delta\tau\right)^2/2\sigma^2\right]}$, where $\sigma$ represents the coherence time of the SPDC field. $\Delta\phi$ contains the phase disturbances arising from mechanical vibrations or airflow disturbances, \textit{etc}., which have significant effects in phase-shifted holography. In general, the entangled photons have a short coherence time $\sigma$, which makes the hologram with high visibility accessible only with an extremely small $\Delta\tau$, where the visibility is $V=[(r_pr_sr_i)/(r_p^2r_s^2)]\gamma(\Delta\tau)$ \cite{gilaberte2021video}. In addition, it is worth noting that the spatial resolution of QHUP is determined by the transverse momentum correlation between the two entangled photons, which is intrinsically limited by the crystal’s aperture and pump beam size. It renders QHUP to display a lower spatial resolution compared with classical holography.
\par

To mitigate these deficiencies on QHUP, we design a convolution-based quantum holography with undetected photons network (QHUPnet) (Fig. \ref{fig1}c). Through learning a series of object-hologram image pairs, QHUPnet can discern the relationship between objects and holograms, such that the one-shot holography can be achieved even with the deteriorated holograms of QHUP. Specifically, QHUPnet contains two stages of extracting abstract features and reconstructing object images from these features \cite{lecun2015deep}. In the first stage, the redundant and irrelevant information, such as interference pattern shifts from the phase disturbance, and random noise from the light source and detectors, are progressively removed by multilayer convolutional and pooling operations, obtaining the abstract features. In the second stage, the learning/training process autonomously models the correspondence from the abstract features to the object images, enabling the one-shot holographic imaging with low interference visibility under strong phase disturbances. Furthermore, depending on the unveiled correlation among pixels in the learning/inference process of QHUPnet, the spatial resolution of QHUP can be improved.\par

QHUPnet consists of atrous spatial pyramid pooling (ASPP), concatenate convolution (Cat. Conv.) and dense block. Convolution is a weights-sharing operation with emphasis on local features in the original patterns. However, the local feature embedded in the holograms undergoes gradual degradation and hence evolves to a nonlocal one due to diffraction. To improve the capacity of processing these nonlocal features, ASPP is implemented in the first stage to facilitate multi-dimensional feature extraction \cite{chen2017deeplab}. Subsequently, for the purpose of enhancing the fidelity of the reconstructed images, the dense block with features reuse and fusion is applied to perform pixel-wise regression from abstract features to object images \cite{huang2017densely}. Notably, the Cat. Conv. module is designed to reduce information loss by providing a path of information flow between feature extraction and image reconstruction. On account of such a delicate architecture, QHUPnet offers a robust and effective solution for reinforcing the imaging performance of QHUP (see Fig. S2 for architecture details).\par 

\begin{figure}[h]%
\centering
\includegraphics[width=1\textwidth]{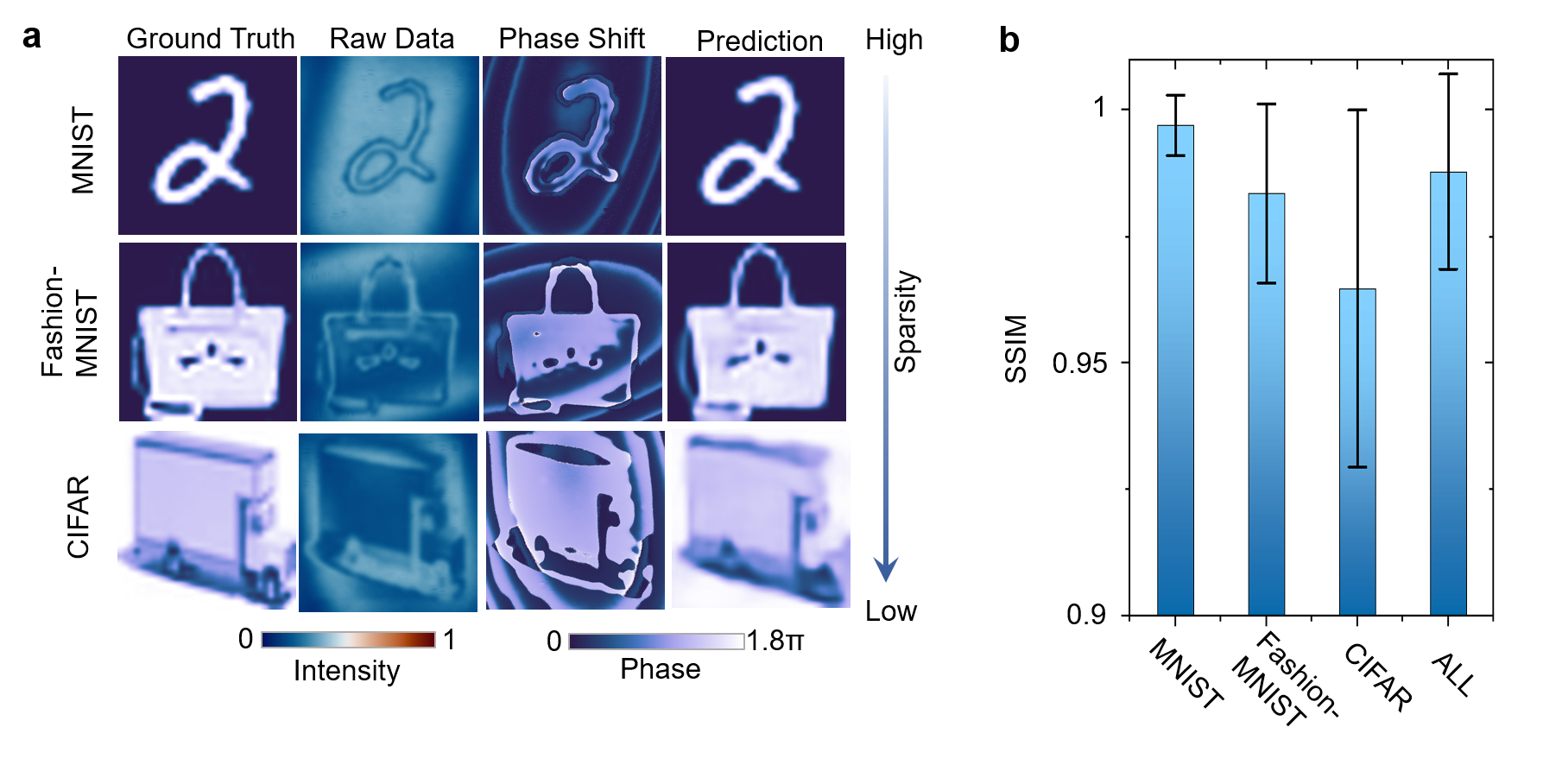}
\caption{\textbf{Performance of DL-QHUP.} (a) Randomly selected ground truth, the corresponding holograms, phase imaging by phase-shifted method, and the prediction of QHUPnet. (b) Performance evaluation by structural similarity index measurement (SSIM) \cite{wang2004image} for MINIST, Fashion-MINIST, and CIFAR, respectively, where the sparsity is decreased. ALL represents the mixture of three datasets. Although QHUPnet is trained with large size dataset, DL-QHUP allows accurate phase prediction with small size data, making it more accessible in practical scenarios (see Fig. S3 for quantitative results). The number of testing images is 15000. Error bar: $\pm$ standard deviation.}
\label{fig2}
\end{figure}

Once QHUPnet is built, DL-QHUP can acquire the phase image through the one-shot hologram where the phase information is thoroughly lost in the intensity measurements. Through the training process, the neural network not only learns the mapping between hologram-object image pairs but also establishes the relationship among these pairs, which ensures the uniqueness of solution of the phase and thus finds the one-to-one correspondence between the holograms and objects. However, the mismatch between the training and testing datasets, which generally occurs for analogue training data and physical testing data, can deteriorate the performance of image reconstruction given by QHUPnet. Therefore, we use hybrid training in QHUPnet to enhance the generalization across three diverse datasets with varying sparsity. Upon completion of the training process, QHUPnet can finally output high-fidelity images from the one-shot hologram. To demonstrate the advantages of QHUPnet compared to conventional phase-shifted methods, the results of QHUPnet on such three distinct datasets with different sparsity, named MNIST, Fashion-MNIST and CIFAR, are shown in Fig. \ref{fig2}.
\par

\begin{figure}[htb]%
\centering
\includegraphics[width=0.94\textwidth]{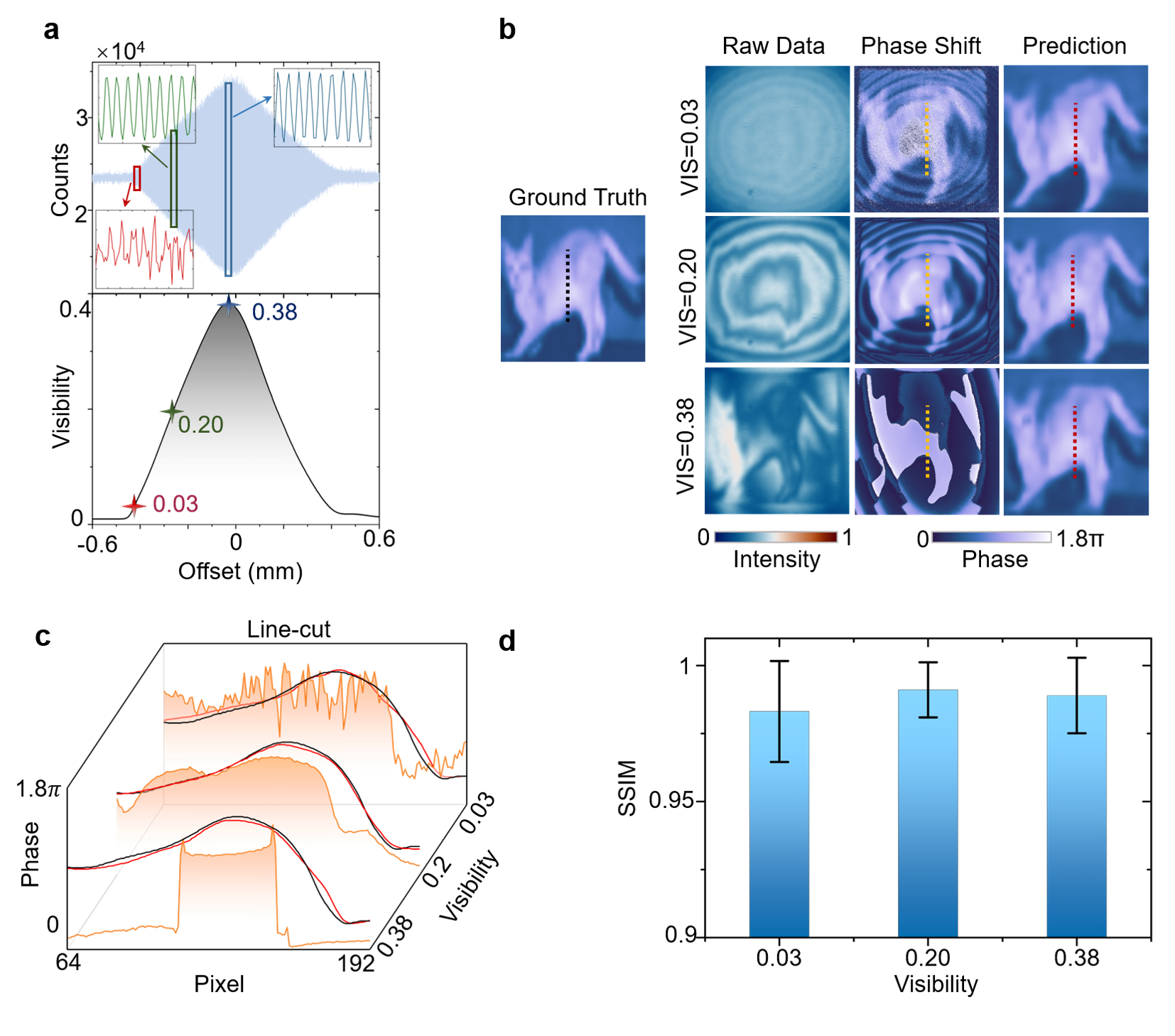}
\caption{\textbf{Performance of DL-QHUP with different interference visibility.} (a) The blue shadow in the top panel indicates that the intensity oscillates with offset distance between signal and idler beams at the central pixel of holograms. The black curve in the bottom panel shows the corresponding average visibility. Inset: interference fringes with three different visibilities (the ranges of vertical axis are 0.2 (red), 1.2 (green) and 2.1 (blue)). (b) Randomly selected results with different visibility. The line-cuts marked by dashed lines are drawn and depicted as the waterfall graphs in (c), where the predictions of DL-QHUP (red) are smooth and consistent with the ground truth (black). The results of phase-shifted method are substantially deviate from the ground truth. (d) The statistics of DL-QHUP with different visibilities. DL-QHUP shows a distinguished capability in information extraction for holograms with different visibilities. Error bar: $\pm$ standard deviation. The number of images is 15000.}\label{fig3}
\end{figure}

For the phase-shifted-based holography, the amplitude noise in one-shot intensity measurements and the phase disturbances accumulated in multi-shot operations can result in seriously negative effects \cite{dainty2013laser}, leading to a large inaccuracy in phase extraction from holograms. Such an inaccuracy severely deteriorates the results in the high- and low-visibility regions (see Fig. \ref{fig3}(a)). Specifically, in the low-visibility region, due to the weakly embedded object features in the hologram, the reconstructed image has a definite contour but with a strong background and intensity fluctuation (see the first row of Fig. \ref{fig3}(b)). On the other hand, as shown in the third row of Fig. \ref{fig3}(b), the phase disturbance arising from the distance variation of idler and signal paths surpasses the intensity noise in one-shot holograms and results in a failed reconstruction of the object image. Remarkably, such two negative effects can be balanced in the intermediate visibility region, and thus enables DL-QHUP to output distinct images as shown in the middle row of Fig. \ref{fig3}(b).\par

In QHUP, the visibility of hologram depends on the distance difference between the signal and idler photons (Fig. \ref{fig3}(a)) and reaches the maximum only at the Fourier plane of the parabolic mirror (shown in Fig. S1). When dealing with real samples, it becomes almost impractical to adjust the optical delay between two beams to ensure the high visibility across the whole imaging region due to the thickness fluctuation, lowering the imaging quality across different regions of the interference (characterized by visibility). Importantly, such an issue can be autonomously compensated by deep learning, and therefore achieves reliable imaging results regardless of the visibility, as demonstrated in Fig. \ref{fig3}(b-d).\par

\begin{figure}[h]%
\centering
\includegraphics[width=0.94\textwidth]{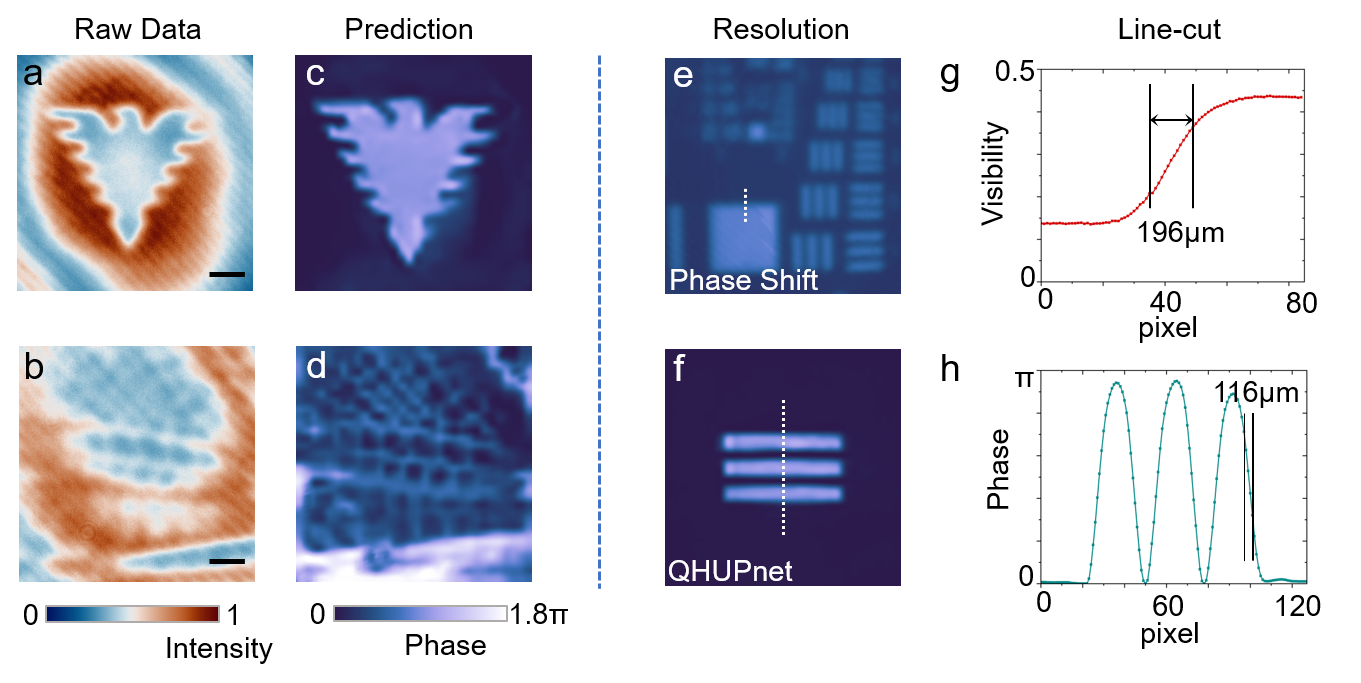}
\caption{\textbf{Imaging for real samples and  spatial resolution improvement.} (a-b) The holograms with “Qiushi” eagle and dragonfly wing. (c-d) The reconstructed images of (a) and (b), respectively. (e) The resolution target image by phase-shifted method. (f) The reconstructed image of an artificial target by DL-QHUP. (g-h) The line-cut highlighted by the white-dashed lines in (e) and (f), respectively. The spatial resolution is fitted by the pixels where the visibility or phase drops from 76\% to 24\% of the maximum, as defined in Ref.~\cite{fuenzalida2022resolution}. The scale bar represents 500 $\mu$m.} \label{fig4}
\end{figure}

The dataset of holograms for training and testing is generated by the virtual objects on a spatial light modulator (SLM). Deep learning is able to find out the similarity in morphological characteristics between training and testing datasets when they are generated from the same virtual objects. For real samples, however, such a requirement for dataset cannot be satisfied and thus leads to a degraded imaging quality due to the inevitable discrepancy between real samples and virtual objects. Nevertheless, the hybridized training of QHUPnet allows to facilitate imaging for real samples. The SLM is replaced by a dragonfly wing and artificial glass-plate with “Qiushi” eagle to test the QHUPnet (see Fig. \ref{fig4}(a-d)).\par

Another significant advantage of DL-QHUP lies in the improvement of the spatial resolution. Recently, numerous studies have unveiled that deep learning can enhance the resolution by learning pixel-level associations across the entire images \cite{wang2020deep,wang2019deep,ouyang2018deep}. QHUP’s resolution is determined by the momentum correlation between the entangled photons \cite{fuenzalida2022resolution}, approximately $196~\rm{\mu m}$ for our configuration through measuring a resolution target (see Fig. \ref{fig4}(e) and (g)), which agrees with the theoretical value $222~\rm{\mu m}$. DL-QHUP displays a capacity in improving the spatial resolution up to $116~\rm{\mu m}$ owing to the appendant nature of QHUPnet through learning massive data pairs without meticulous selection of training data or specific design and labeling of setups and samples (see Fig. \ref{fig4}(f) and (h)).\par

\section*{Discussion}
DL-QHUP is capable of recording the complex amplitude of objects by training QHUPnet with the dataset containing simultaneous modulation in amplitude and phase. Traditionally, such a modulation can be implemented by computational holography \cite{ulusoy2011full,goorden2014superpixel} or utilizing multiple modulators working at either amplitude-only or phase-only mode \cite{neto1996full}. Requiring diffraction or multiple modulations at various spatial positions, these methods generally lead to a large loss in photons, rendering incompatibility with QHUP. Fortunately, entangled photons provide a feasible approach to modulate complex amplitude. For the sake of clarity, considering the complex amplitude of the object is $r_o \exp{\left(i\varphi_o\right)}$, the distribution of intensity at the detector is,
\begin{equation}
\begin{aligned}
I_s & \propto r_p^2+r_s^2+2 r_p r_s \hat{r}_i \cos \left(\varphi_s+\hat{\varphi}_i-\varphi_p+\Delta \phi\right) \\
%& =r_p^2+r_s^2+2 r_p r_s r_i r_o \cos \left[\varphi_s+\left(\varphi_i+\varphi_o\right)-\varphi_p+\Delta \phi\right] \\
& =r_p^2+r_s^2+2 r_p r_s \hat{r}_i \cos \left(\hat{\varphi}_s+\varphi_i-\varphi_p+\Delta \phi\right)
\end{aligned}
\label{eq2},
\end{equation}
where $\hat{\varphi}_{s/i}=\varphi_{s/i}+\varphi_o$ and $\hat{r}_{s/i}=r_{s/i}{\cdot}r_o$ (assume $\gamma(\Delta\tau)=1$). In Eq. (\ref{eq2}), the phases of the signal and idler photons have equal roles in the interference. However, only $r_s^2$ affects the intensity distribution of holograms while $r_i^2$ is absent.  Thus, simultaneously modulating the phase of signal photons and amplitude of idler photons is equivalent to modulating the complex amplitude of idler photons. As the results shown in Fig. S4, this feature enables complex amplitude imaging by DL-QHUP and can be generalized to more sophisticated applications. \par 

Based on the convolutional neural network, DL-QHUP can achieve higher-fidelity images from holograms with more local information. Typically, more local pattern can be produced by averaging the hologram with more spectral components. However, SLM generally works at single wavelength, hence the modulation of a wide spectrum cannot be performed to establish the training dataset. To balance the degree of local information and the SLM’s modulation eﬀiciency, a modest bandpass spectral filter is inserted in the path of signal photons to equivalently narrow the spectrum of idler photons within the working range of SLM, while the idler photons retain multiple spectral components (see Fig. S5). To validate the influence of spectral components on imaging, we feed the holograms generated by QHUP and by classical Michelson-type interferometers with narrow-band idler photons and laser, respectively, into QHUPnet for image predictions. As shown in Fig. S6, under identical conditions, QHUPnet exhibits optimal performance in QHUP, followed by narrow-band idler photons and laser interferometers, implying that more wavelength components ensure QHUP a better image reconstruction.\par

Compared to the classical holography suffering from the scattered light with the same wavelength as the illumination \cite{qian2023quantum,defienne2021polarization}, QHUP ensures an unaffected hologram without additional handling of scattered light owing to the disparity of wavelengths between the illuminating and detected light. As shown in Fig. S7, LED and laser sources are placed behind of the object for demonstrating a robustness to bright backgrounds. Fig. S8 shows an excellent imaging performance of DL-QHUP in a wide range of working distance, further extending broad applications in practical scenarios.

\section*{Conclusions}
In summary, we have established for the first time that deep learning can mitigate the limitations existing in conventional QHUP through its capability in handling complex data. By harnessing the quantum entanglement of photons, QHUP can simultaneously acquire the amplitude and phase images of objects at flexible wavelengths without being affected by stray light and/or light background. Instead of time-consuming joint measurements (several hours), QHUP empowers rapid image capture within milliseconds through intensity interference control and can further be accelerated by deep learning to single-shot imaging. Notably, the integration of deep learning technology proves pivotal in resolving phase disruptions inherent in QHUP's intricate three-beam setup, significantly enhancing its adaptability across diverse applications. Even in scenarios of minimal interference visibility, deep learning algorithms excel in extracting essential holographic features and reconstructing high-fidelity images with remarkable precision.

Furthermore, the synergy between deep learning and quantum optics enables a leap in spatial resolution enhancement by leveraging pixel correlations. This transformative advancement is exemplified through successful imaging of real-world samples, showcasing the profound impact of this convergence on imaging capabilities. This work not only showcases the comprehensive advantages of uniting deep learning and quantum optics but also maps a path towards a multitude of practical applications. From quantum light detection and ranging \cite{qian2023quantum} to biomedical diagnostics \cite{balasubramani2021holographic} and environmental monitoring \cite{wu2018lensless}, the potential implications of this technology span across diverse fields, particularly in the realms of infrared and terahertz imaging \cite{kutas2021quantum,heimbeck2020terahertz,wan2020terahertz}. The synergy of deep learning and quantum optics in QHUP lays the foundation for a quantum leap in imaging technology, promising unprecedented advancements and solutions to complex challenges in various application domains.

\section*{Materials and Methods}

\subsection*{Experimental Setup}
As shown in Fig. S1, a 532 nm pump laser is generated by a fiber laser (Precilaser, YFA-SF-1064-50-CW) and directed into a periodically poled lithium niobate (PPLN; Covesion, MSHG-1064-1.0-2.0) to generate approximately 400 nW entangled photon pairs (signal and idler) through the SPDC process. A parabolic mirror (PM; Thorlabs, MPD169-P01) with a focal length of 150 mm is used to collimate the entangled beams. Subsequently, the beams are spatially separated by two dichroic mirrors (DM; Thorlabs, DMLP650 for DM2 and DMSP1000 for DM3) into the signal, idler, and pump beams. A half-wave plate (HWP) in front of SLM (Hamamatsu, X13139–09) adjusts the polarization of idler beam to match the modulation axis of SLM. The distance difference between the signal and idler beams is controlled by the M2, a movable mirror on a translation stage. The pump, signal, and idler beams are reflected by M1, M2, and SLM, respectively. These beams re-enter the PPLN crystal and initiate the SPDC a second time. The pump beam is separated from the mixed beam with DM1 (Semrock, FF700-SDI01), and eliminated by an optical isolator (OI; LBTEK, ISO532-3-1.5W). Before the signal beam entering the detector, a bandpass filter (BP; Thorlabs, FL905-10) is used to narrow the spectrum of signal beam. An image with a size of $512 \times 512$ pixels and 16-bit grayscale pattern is recorded by a sCMOS camera (Tucsen, Dhyana 400D). Specially, the distances are 170 mm from the PM to SLM, 75 mm from the lens to sCMOS, and the system operates in a non-imaging configuration. When implementing real sample imaging, we replace SLM by real objects. The acquisition time for each hologram is 10 ms.\par

\subsection*{Phase Extraction by the Phase-Shifted Method}
Considering a phase-only object under the illumination of completely coherent beam, the intensity distribution of signal photon is $I_s \left(\varphi_o, m\alpha\right) \sim 1 + \cos{\left(\varphi_o + m\alpha + \varphi_{BG}\right)}$, where $\varphi_{BG}$ is the background phase of the system, $\varphi_o$ represents the phase of the object, and $\alpha=2\pi/N$ is the interval of phase shift with $m = 0,1,2,\dots,N-1$. In our experiment, the number of phase-shifted step is 4, which balances both retrieval quality and exposure number. In order to subtract the influence of background phase, we collect the holograms of background-only by performing phase-shifted method. By eight measurements, the phase of object can be extracted, which follows as,
\begin{equation}
\varphi_o=\arg \left[\frac{\left(I_s\left(\varphi_o, 0\right)-I_s\left(\varphi_o, \pi\right)\right)+i\left(I_s\left(\varphi_o, 3 \pi / 2\right)-I_s\left(\varphi_o, \pi / 2\right)\right)}{(I_s(0,0)-I_s(0, \pi))+i(I_s(0,3 \pi / 2)-I_s(0, \pi / 2))}\right].
\end{equation}
Examples of the data and corresponding results are shown in Fig. S9.

\subsection*{QHUPnet implementation}
\textit{Structure}: upon the input of $512 \times 512$ hologram, a compressed feature tensor of $256 \times 256 \times 32$ (height $\times$ width $\times$ channel) is acquired by Conv. Pool. which consists of $2 \times 2$ convolution, batchnormalization, and Hardswish activation functions. The low-level feature tensor then goes through ASPP module for five times to extract more advanced features with size of $128 \times 128 \times 64$, $64 \times 64 \times 128$, $32 \times 32 \times 256$, $16 \times 16 \times 512$, and $8 \times 8 \times 1024$ blocks, respectively, defined as the encoding process. In this process, to avoid significant loss of information, the Cat. Conv. module is well-designed for feature fusion of two feature tensors with different resolutions of successive layers. The obtained feature tensor is defined as intermedium tensor. In the decoding process, the Cat. Conv. Module is used again for feature fusion of the high-level feature and the intermedium tensors. Subsequently, the resultant feature tensor is decoded and completed by the dense block module, where the spatial size of the feature tensor is doubled and the number of channels is reduced by half. After repeating these processes by four times, the size of feature tensor becomes $256 \times 256 \times 32$, and the decoding process is finished. The feature tensor is compressed to 16 channels by $3 \times 3$ convolution, and then the $1 \times 1$ convolution is used to perform depth-to-space operation to obtain the output with 1 channel, actually the final image. If the phase and amplitude are simultaneously extracted from the single hologram, only two $1 \times 1$ convolutions which share the previous feature tensor are required. We note that the residual-like method is used in each module. See Fig. S2 for the calculation details and hyperparameters of each module.\par 

\textit{Train}: when the network is built, QHUPnet is trained and tested under the PyTorch framework \cite{paszke2019pytorch}, and is performed on the server (24 GB RAM, Xeon E5-1654 v4, 3090 Ti, CUDA 11.6). The mean square error (MSE), Pearson correlation coeﬀicient (PCC), and structural similarity index measure (SSIM) are used to evaluate the similarity between the predicted image and the ground truth. MSE exhibits the capability of easily perceiving low-frequency information (profile) of image and thus is more suitable for sparse target reconstruction. Alternatively, PCC is effective for the non-sparse target reconstruction by capturing the feature with local correlation. Additionally, considering the image perception of human vision, the multi-scale SSIM serves as a term in the loss function to local illumination, contrast, structure of image. To speed up the training and obtain better imaging quality, the complete loss function $\mathcal{L}$ is defined as:
\begin{equation}
\mathcal{L}=4\mathcal{L}_{MSE}+\mathcal{L}_{NPCC}+\mathcal{L}_{MSLL},
\end{equation}
where MSE is a built-in function in PyTorch library while NPCC and MSLL are the forms of the loss function corresponding to PCC and multi-scale SSIM \cite{paszke2019pytorch,zhao2016loss}, respectively. In our experiment, the modified national institute of standards and technology (MNIST) dataset \cite{deng2012mnist}, Fashion-MNIST dataset \cite{xiao2017fashion}, and CIFAR-10 dataset \cite{krizhevsky2009learning} are used as the input images for SLM and the ground truth images for QHUPnet. For each collected dataset, 90\% and 10\% of the dataset are used for training and testing, respectively. Each dataset is fed into the QHUPnet and trained with 15 epochs by the AdamW optimizer. The AdamW minimizes the loss function with an initial learning rate 0.001, weight decay 0.0005, where amsgrad is applied.\par

\section*{Declarations}
\subsection*{Availability of data and materials}
Data underlying the results presented in this paper are not publicly available at this time but may be obtained from the authors upon reasonable request.
\subsection*{Competing interests}
The authors declare that they have no competing interests.
\subsection*{Funding}
The National Natural Science Foundation of China (Grant No. 11934011, 62075194, U21A6006). The National Key Research and Development Program of China (Grant No. 2019YFA0308100, 2023YFB2806000, 2022YFA1204700). The Strategic Priority Research Program of Chinese Academy of Sciences (Grant No. XDB28000000). The Fundamental Research Funds for the Central Universities. The Information Technology Center and State Key Lab of CAD\&CG.
\subsection*{Authors' contributions}
W.F., C.X. and X.X. conceived the idea and designed the experiment. W.F. and G.Q. carried out the experiment, collected data and performed numerical simulation. G.Q. built the optical systems, and W.F. wrote the control programs for devices. Y.W. fabricated the artificial phase plates. W.F., G.Q., X.X. and C.X. analysed data and wrote the manuscript. The project is supervised under D.W.W. All authors discussed the results and revised the manuscript.

\subsection*{Acknowledgements}
Not applicable.
\subsection*{Supplementary information}
Please see Supplementary material for supporting contents including Table 1 and Figs. S1-S9.

\end{document}

\title{Supplementary Material: Deep Learning Enhanced Quantum Holography with Undetected Photons}

\author[1]{\sur{Weiru} \fnm{Fan}} 
\equalcont{These authors contributed equally to this work.}
\author[1]{\sur{Gewei} \fnm{Qian}}
\equalcont{These authors contributed equally to this work.}
\author[2]{\sur{Yutong} \fnm{Wang}} 
\author*[1]{\sur{Chen-Ran} \fnm{Xu}} \email{crxu@zju.edu.cn}
\author[3]{\sur{Ziyang} \fnm{Chen}}
\author[4]{\sur{Xun} \fnm{Liu}}
\author[4]{\sur{Wei} \fnm{Li}}
\author[5]{\sur{Xu} \fnm{Liu}}
\author[2]{\sur{Feng} \fnm{Liu}}
\author*[1]{\sur{Xingqi} \fnm{Xu}} \email{xuxingqi@zju.edu.cn}
\author[1,5,6]{\sur{Da-Wei} \fnm{Wang}}
\author*[7]{\sur{Vladislav V.} \fnm{Yakovlev}}\email{yakovlev@tamu.edu}

\affil[1]{\orgdiv{Zhejiang Province Key Laboratory of Quantum Technology and Device, School of Physics, and State Key Laboratory for Extreme Photonics and Instrumentation}, \orgname{Zhejiang University}, \orgaddress{\city{Hangzhou}, \postcode{310027}, \state{Zhejiang Province}, \country{China}}}
\affil[2]{\orgdiv{College of Information Science and Electronic Engineering}, \orgname{Zhejiang University}, \orgaddress{\city{Hangzhou}, \postcode{310027}, \state{Zhejiang Province}, \country{China}}}
\affil[3]{\orgdiv{College of Information Science and Engineering, Fujian Key Laboratory of Light Propagation and Transformation}, \orgname{Huaqiao University}, \orgaddress{\city{Xiamen}, \postcode{361021}, \state{Fujian Province}, \country{China}}}
\affil[4]{\orgdiv{Beijing Institute of Space and Electricity}, \orgname{China Academy of Space Technology}, \orgaddress{\city{Beijing}, \postcode{100094}, \country{China}}}
\affil[5]{\orgdiv{College of Optical Science and Engineering}, \orgname{Zhejiang University}, \orgaddress{\city{Hangzhou}, \postcode{310027}, \state{Zhejiang Province}, \country{China}}}
\affil[6]{\orgdiv{Hefei National Laboratory}, \orgaddress{\city{Hefei}, \postcode{230088}, \state{Anhui province}, \country{China}}}
\affil[7]{\orgdiv{Department of Biomedical Engineering}, \orgname{Texas A\&M University}, \orgaddress{\city{College Station}, \postcode{77843}, \state{TX}, \country{USA}}}

\maketitle

\begin{table}[t!]
\renewcommand\arraystretch{1.4}
\caption{\textbf{Comparison between different holography modalities}}\label{tab1}
\begin{tabular}{ccccccc}
\rowcolor{black!12}
& \makecell*[c]{Single\\shot} & \makecell*[c]{Model\\free} & \makecell*[c]{Reference\\free} & \makecell*[c]{Joint-measurement \\ free} & \makecell*[c]{Background\\resilience} & \makecell*[c]{Wavelength\\conversion } \\

\rowcolor{black!6}
\textbf{Classical way}  & & & & & &  \\

TIE \cite{zuo2020transport} & \checkmark & & \checkmark   & \checkmark &  &    \\

Off-axis \cite{cuche2000spatial} & \checkmark &  &  & \checkmark  &  & \\
\makecell*[c]{Phase shift \cite{schnars2015digital}} &  & \checkmark &  & \checkmark  & & \\ 
\makecell*[c]{Deep learning \cite{rivenson2018phase}} & \checkmark  & \checkmark  & \checkmark  & \checkmark &  &  \\

%\rowcolor[rgb]{.99,.91,.95}
\rowcolor{black!6}
\textbf{Quantum way}   & & & & &  &  \\

\makecell*[c]{Polarization\\entanglement\cite{defienne2021polarization}} &  &  &  &  & \checkmark  &    \\

\makecell*[c]{QHUP with\\phase shift \cite{topfer2022quantum}}  & & \checkmark  &  & \checkmark & \checkmark &\checkmark  \\
\makecell*[c]{QHUP with \\off-axis \cite{leon2024off} }&\checkmark &  &  & \checkmark & \checkmark & \checkmark  \\
\makecell*[c]{\textcolor{magenta}{DL-QHUP } \\ \textcolor{magenta}{(Ours)}} & \checkmark & \checkmark & \checkmark& \checkmark&\checkmark & \checkmark \\

\hline
\end{tabular}
\end{table}

\clearpage

\begin{figure}[H]
\centering
\includegraphics[width=1\textwidth]{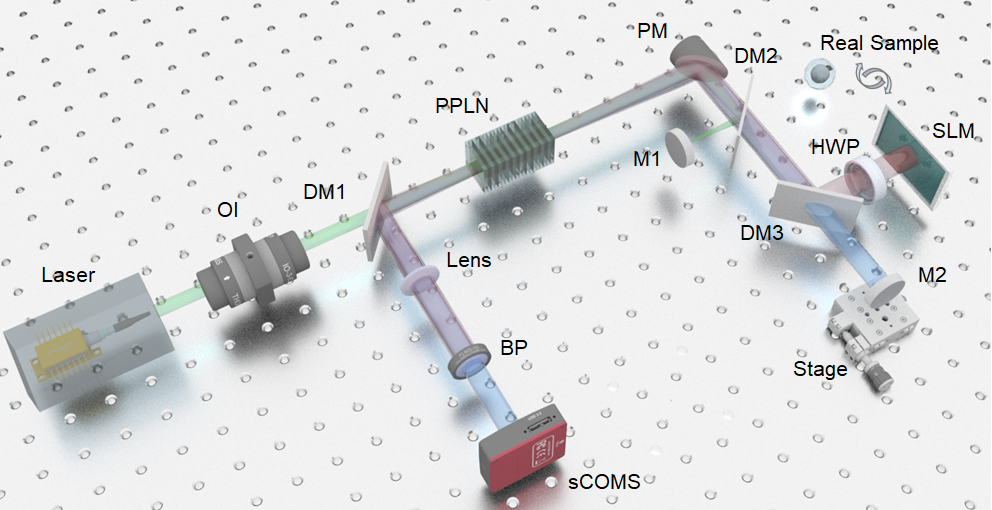}
\caption{\textbf{Experimental setup.} The nonlinear crystal (periodically poled lithium niobate; PPLN) is pumped by continuous wave (green) emitted by the laser to produce entangled signal (blue) and idler (red) beams. Afterwards, the pump beam is separated by the dichroic mirror (DM2), while the signal and the idler beams are separated by DM3. The spatial light modulator (SLM) or objects are placed into idler path to acquire training dataset and holograms with real samples. The bandpass filter (BP) is used to select signal photons with certain wavelength, and filter out the idler photons. The off-axis parabolic mirror and lenses are used to collimate and collect lights, and the SLM or objects cannot be imaged on scientific complementary metal-oxide semiconductor (sCMOS) camera. M1 and M2 are mirrors; HWP is half-wave plate; OI is optical isolator.} \label{figed1}
\end{figure}
\clearpage

\begin{figure}[H]%
\centering
\includegraphics[width=0.8\textwidth]{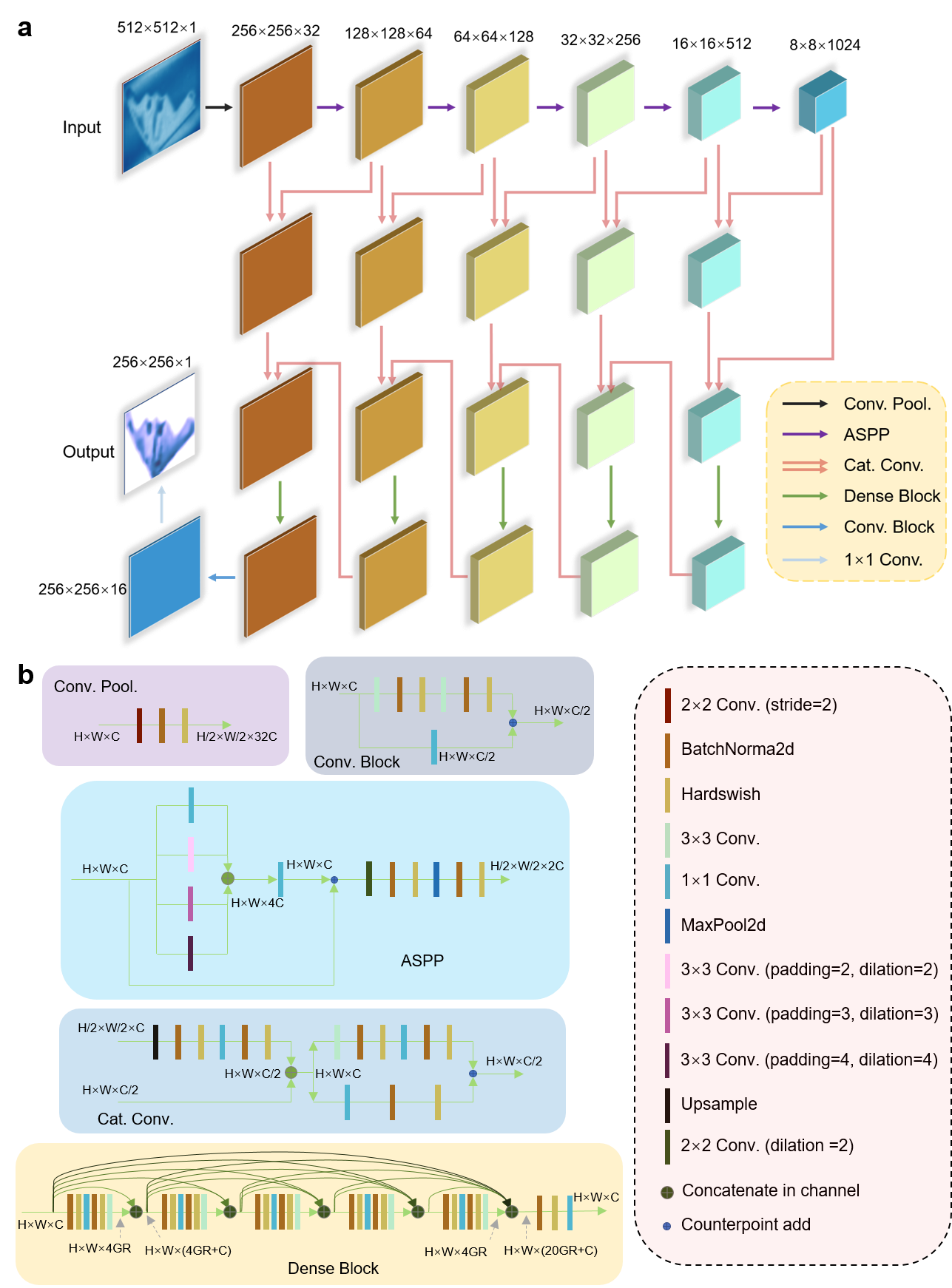}
\caption{(a) The architecture and workflow of QHUPnet. The one-shot hologram is fed into QHUPnet as sole input while the output can be adjusted by final $1 \times 1$ convolution layer according to expectant image dimension. The block/layer with the same color and volume indicates the same size of the feature map. (b) The details for each module in QHUPnet. n $\times$ n Conv. represents the convolution operation with n $\times$ n kernel. GR is the growth rate and equal to (32, 16, 16, 16, 16) in order. H, W and C are the height, width and channel of the image or feature map, respectively.} \label{figed2}
\end{figure}
\clearpage
% \newpage

\begin{figure}[H]%
\centering
\includegraphics[width=1\textwidth]{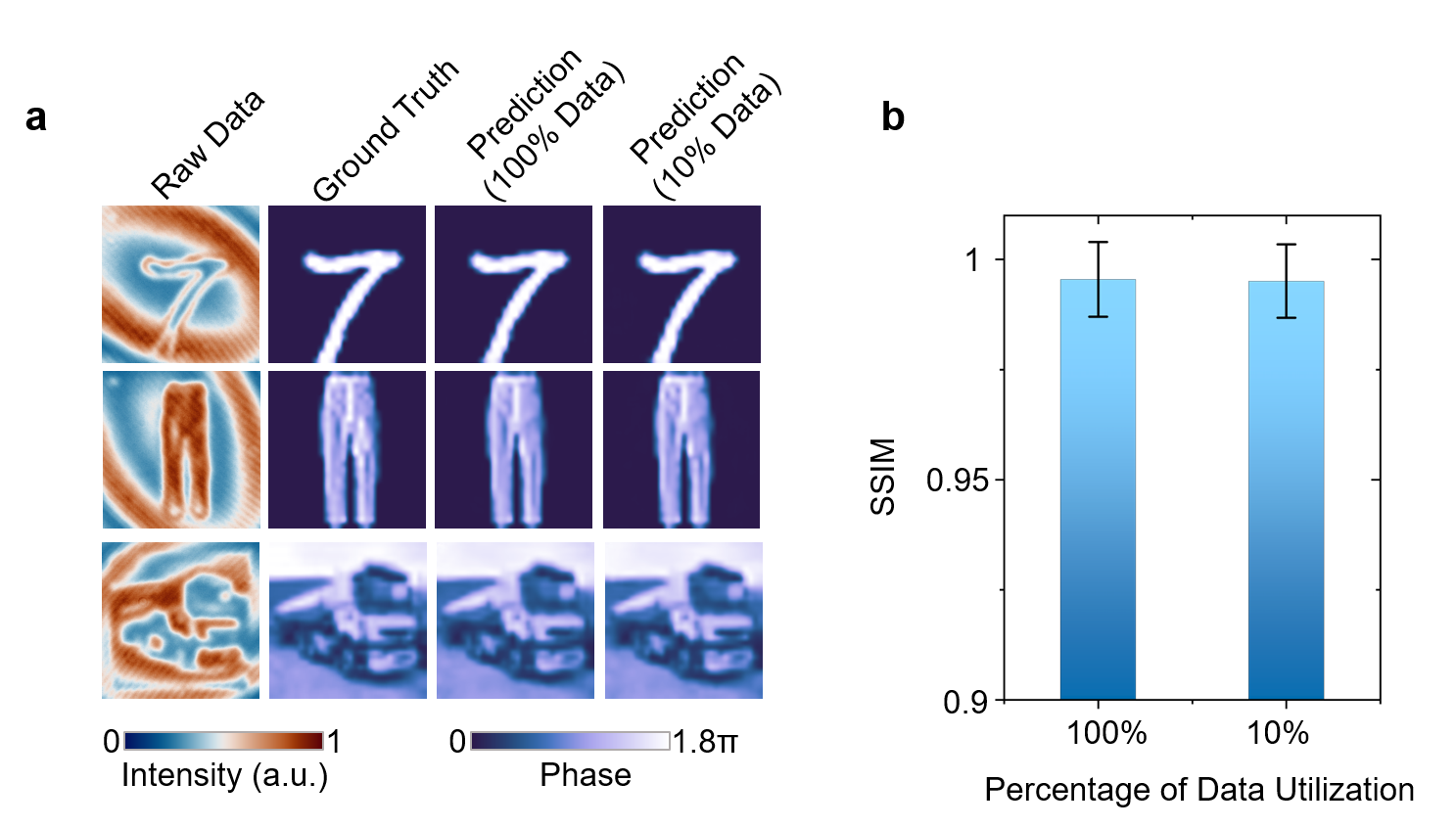}
\caption{\textbf{The performance test of QHUPnet with different sizes of training datasets.} QHUPnet holds excellent performance until the the number of training data is reduced to 10\%. The size of complete dataset is 150,000 (90\% is used to train QHUPnet). Error bar: $\pm$ standard deviation.} \label{figed3}
\end{figure}
\clearpage
% \newpage

\begin{figure}[H]%
\centering
\includegraphics[width=1\textwidth]{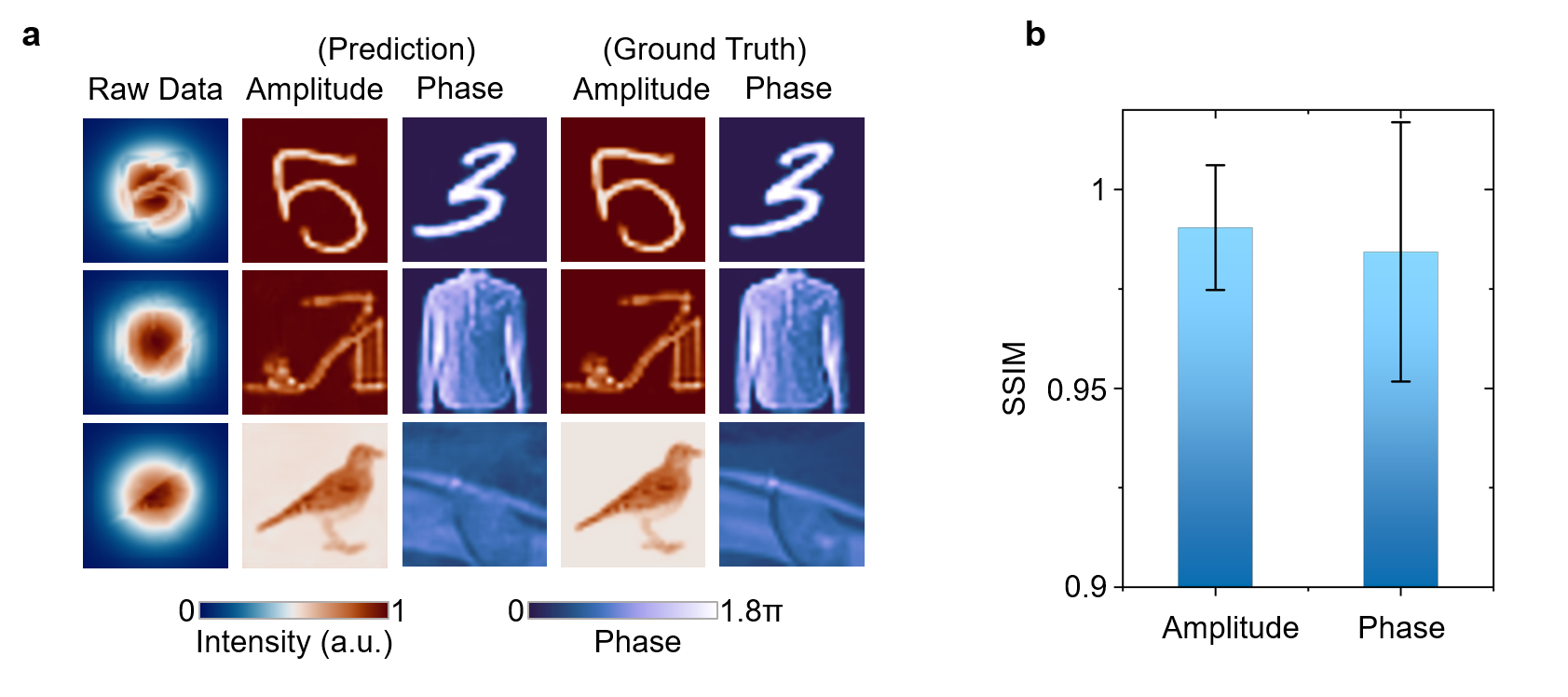}
\caption{\textbf{Imaging complex-amplitude objects by DL-QHUP.} (a) Randomly selected ground truth, the corresponding holograms and the predictions of QHUPnet. (b) Performance evaluation of testing dataset by SSIM. Error bar: $\pm$ standard deviation. The number of images is 15000.} \label{figed5}
\end{figure}
\clearpage
% \newpage

\begin{figure}[H]%
\centering
\includegraphics[width=1\textwidth]{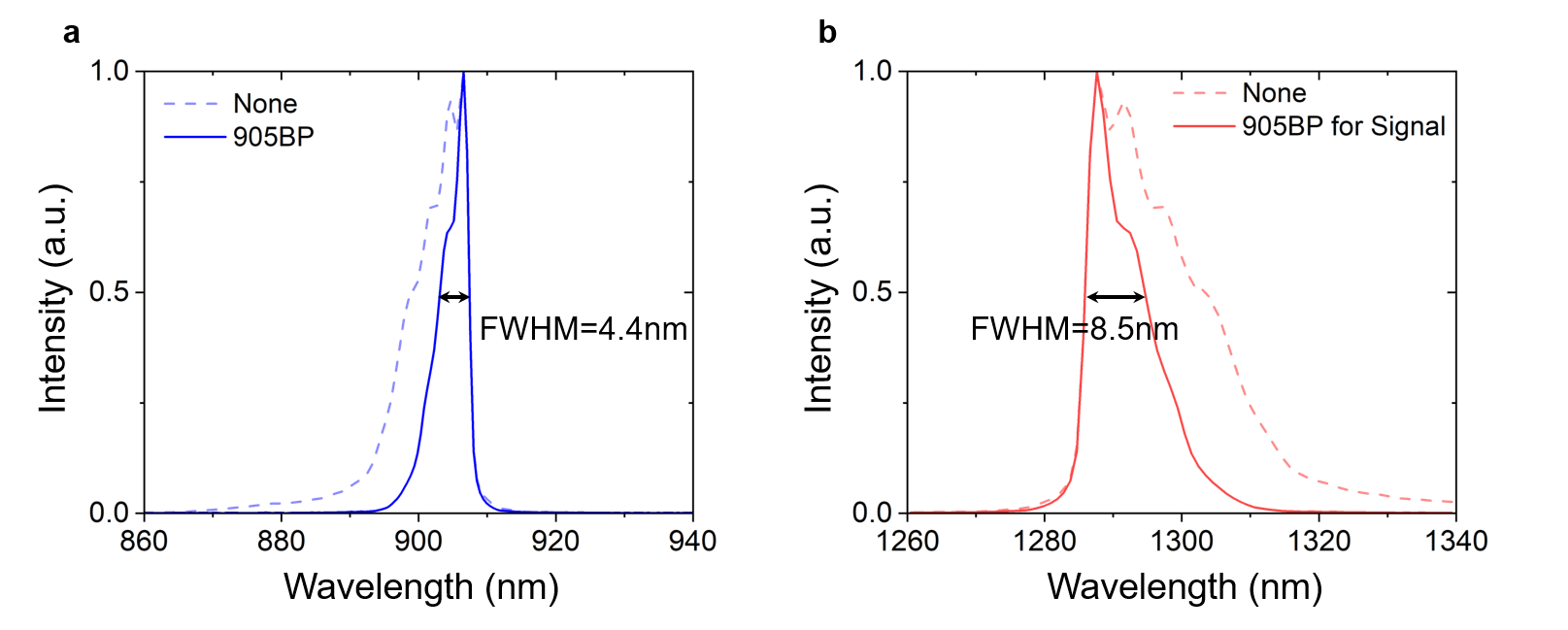}
\caption{(a) and (b) are spectra of signal and idler photons, respectively. The blue and red dashed lines denote the original spectra while the blue and red solid lines represent the narrowed spectra of signal and idler photons with a 905 nm bandpass filter inserted in signal path. The full width at half maximum (FWHM) of signal photons is twice the one of idler photons with the bandpass filter.} \label{figed6}
\end{figure}
\clearpage
% \newpage

\begin{figure}[H]%
\centering
\includegraphics[width=1\textwidth]{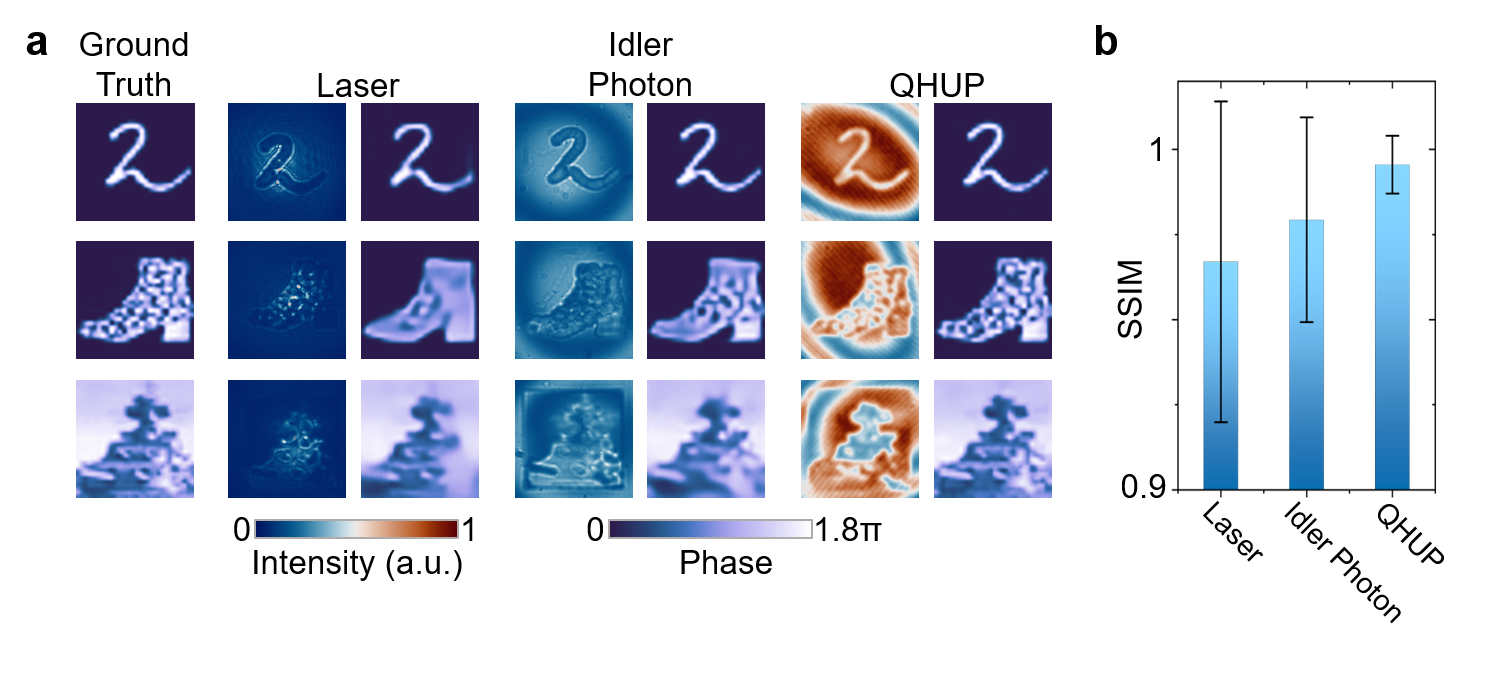}
\caption{\textbf{Imaging performance of different interferometers.} (a) Randomly selected ground truth, the corresponding holograms and the predictions of QHUPnet for interferometers with laser, idler photons and QHUP, respectively. (b) Performance evaluation of the testing dataset by SSIM. The predicted results from QHUP outperform other two interferometers. The data obtained by laser interferometer exhibit large variations, resulting from the significant non-local diffraction and crosstalk among pixels of images. Error bar: $\pm$ standard deviation. The number of images is 15000.} \label{figed7}
\end{figure}
\clearpage
% \newpage

\begin{figure}[H]%
\centering
\includegraphics[width=1\textwidth]{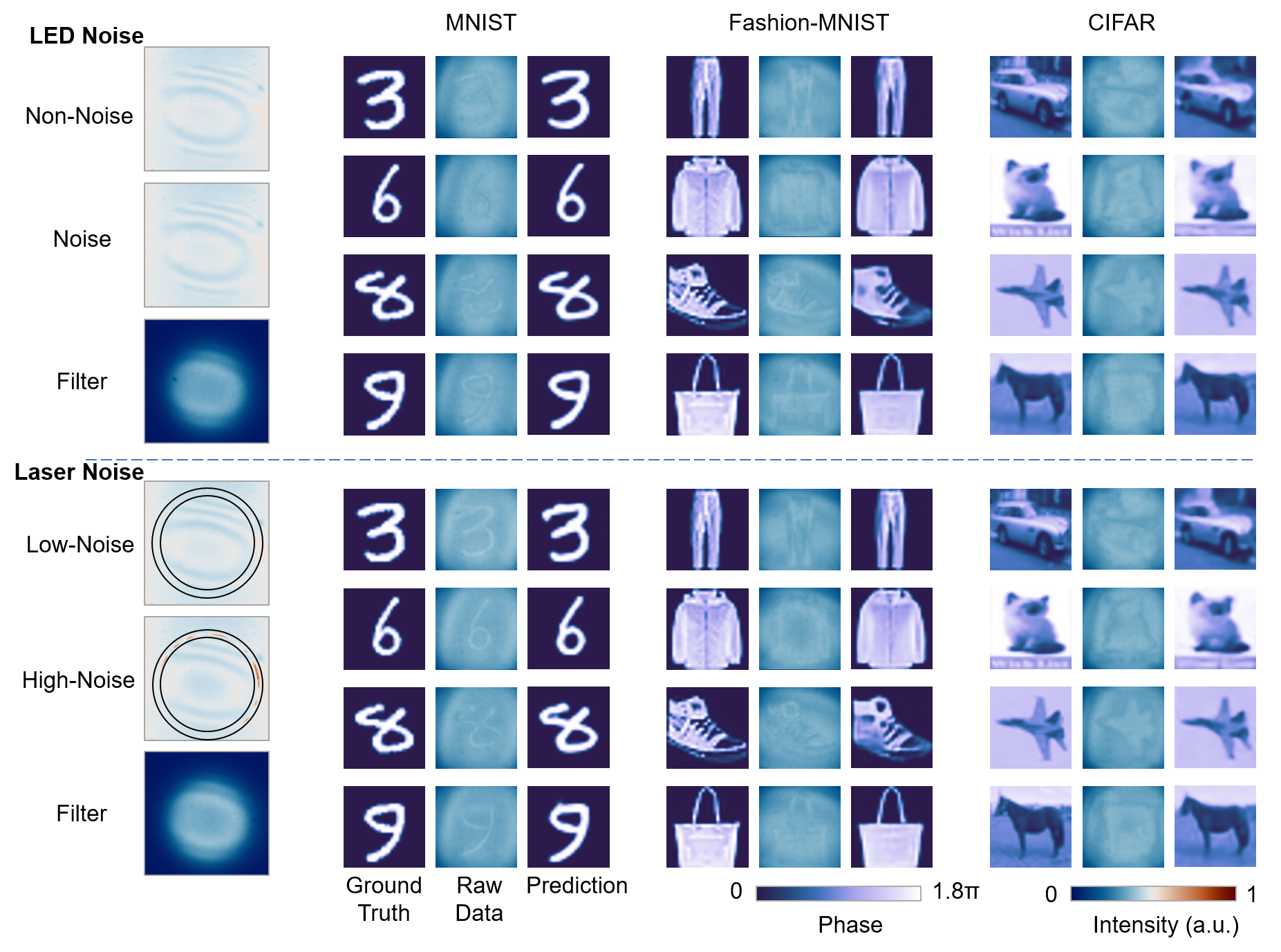}
\caption{\textbf{DL-QHUP performance with background noise.} The first column is the raw data with and without the 905 nm bandpass filter for signal photons. The solid ring is the affected area arising from stimulated emission. 
We set the power of laser in "low-noise" case equal to the power of idler photons, while adjust the powers of laser and LED in the cases of "high-noise" and "LED noise" three orders larger, respectively.} \label{figed8}
\end{figure}
\clearpage
% \newpage

\begin{figure}[H]%
\centering
\includegraphics[width=1\textwidth]{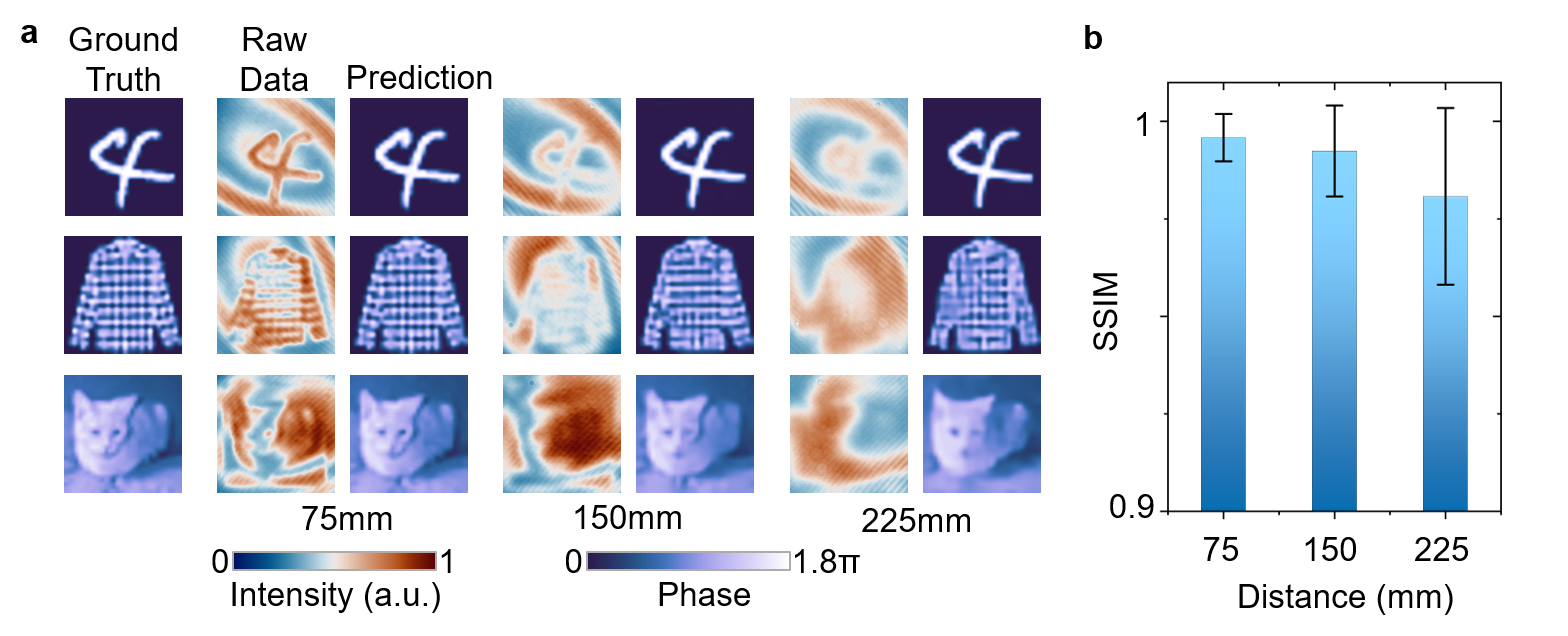}
\caption{\textbf{DL-QHUP imaging at different detection distances.} (a) Randomly selected ground truth, the corresponding holograms and the predictions of QHUPnet. (b) Performance evaluation of the testing dataset by SSIM. Although the detection distances deviate from the imaging plane, DL-QHUP can work effectively. For comparison, the data with different detection distances are trained in the same initial weights and epoch. The imaging quality can be improved by increasing the training steps. Error bar: $\pm$ standard deviation. The number of images is 15000.} \label{figed9}
\end{figure}
\clearpage
% \newpage

\begin{figure}[h]%
\centering
\includegraphics[width=0.7\textwidth]{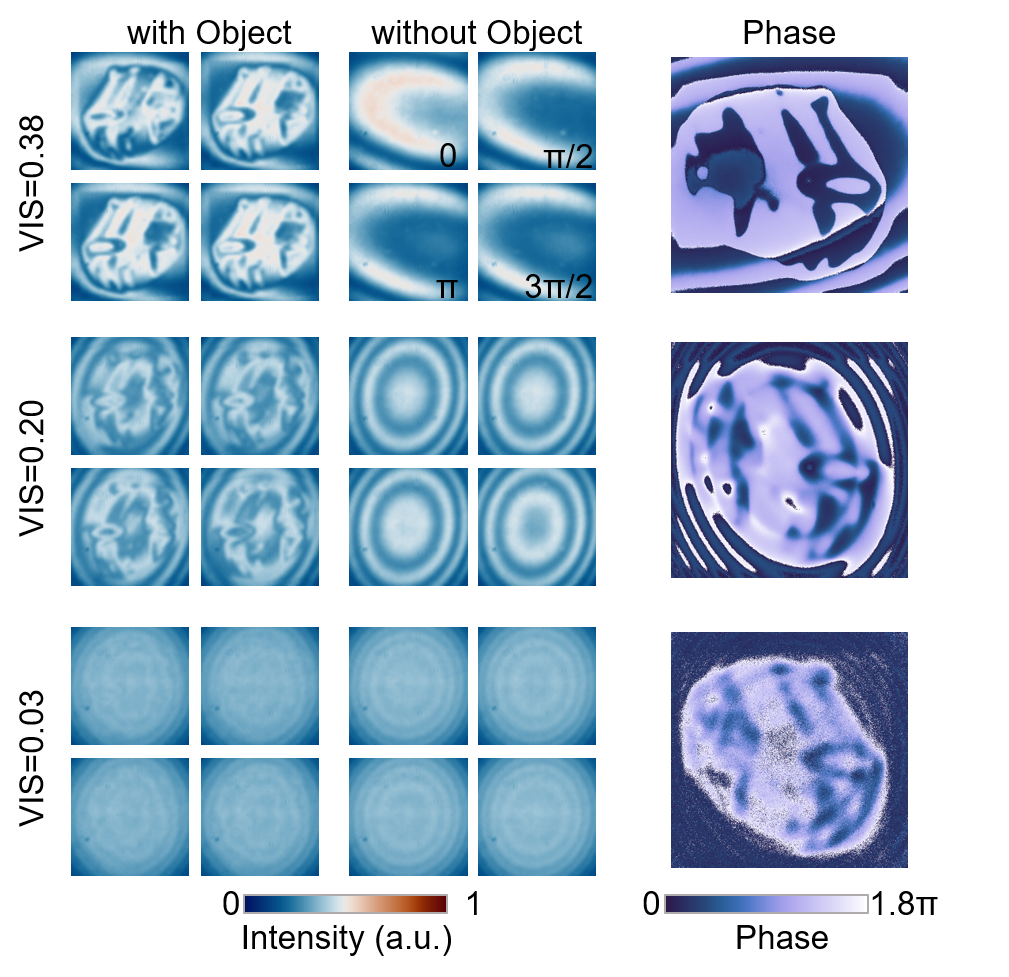}
\caption{\textbf{A set of holograms for the phase retrieval by four-step phase-shifted method with different visibility.} The holograms with the object (the first and second columns), the reference holograms without the object (the third and fourth columns) and the final results of phase imaging (the fifth column) are drawn, respectively.} 
\label{figed10}
\end{figure}

\newpage